# About the Mechanism of Geyser Eruption

Andrei Nechayev

*Moscow State University, Geographic Department,
Laboratory of Renewable Energy Sources*

## Abstract

*Essentially new physical mechanism of geyser eruption based on instability in "water-vapor" system is proposed. Necessary and sufficient conditions of eruptions are received. For group of Kamchatka geysers a good accordance of theoretical model with empirical observations is shown.*

## Introduction

Looking at the map of world geyser location (www.johnstonsarchive.net/geysers), one can notice that this natural phenomenon is found practically on all continents, on many islands, but everywhere in rather limited quantities. Only two places of geyser's residence considerably differ from the others. It is Yellowstone National Park in the USA and the Valley of Geysers on Kamchatka (Russia). The Yellowstone counts more than 500 geysers and in the Valley of Geysers there are nearby 200. In Yellowstone geyser location occupy whole "country" with the area more than thousand sq.km. In the Valley of Geysers most of thermal springs are restricted on a territory in 3-4 sq.km. The concentration of geysers in the Valley and their variety are unique. Unfortunately, after the catastrophic mudflow on June, 3rd, 2007 some beautiful geysers of the Valley were lost irrevocably.

Geyser as the natural phenomenon drew to itself the attention of researchers for a long time (Ustinova, 1955; Allen, Day, 1935). Many models of its structure were offered ( Lloyd, 1975; Droznin, 1982; Steinberg, etc., 1984). They well described the main physical properties of geysers but any of models didn't receive still the general approval and any mechanism couldn't explain all known variety of geysers and features of their behavior. One reason of it is the absence of empirical data about real structure of a geyser. We propose a detailed description of new physical mechanism of geyser eruption based on the well-known "chamber" model (Lyell, 1830) where the instability mechanism starts under definite conditions in the water-vapor system. The idea of this mechanism was proposed in Reykjavik (Nechayev, 2008).

## 2. Theoretical model and the criterion of existence of a geyser.

Let's consider the so-called "chamber" model (fig. 1). A geyser has a vertical channel connecting surface of the earth and the lateral subterranean chamber. In this chamber a boiling and steam formation occur. Let's call it "boiler" to distinguish this model from others. The channel depth is $H$, the section is $S$, the boiler volume is $V_b$. In the lower part of geyser structure the temperature of surrounding rocks exceeds boiling temperature at the corresponding pressure. A significant condition is that the arch of a boiler is impenetrable for vapor. Steam can leave structure only via the geyser channel.

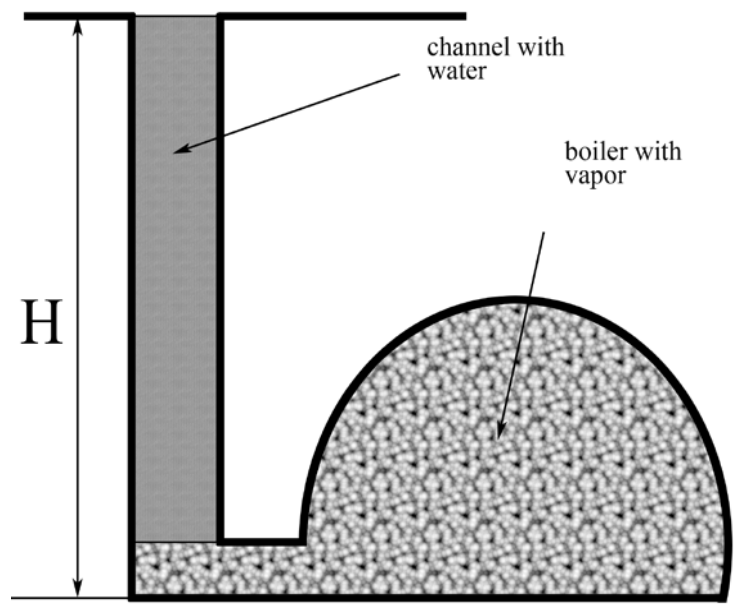

Fig. 1 Chamber model of a geyser. $H$ is the depth of a channel, $S$ is the section. The boiler with the volume $V_b$ is filled with steam which starts to penetrate into the canal.

Let's consider 4 phases of work of a traditional geyser: water filling, water overflow, eruption and steaming. An initial condition: the structure is empty. Then the channel and the boiler are filled with water. The origin of this water and its temperature for the instability mechanism is unimportant. It can be hot or cold, from underground sources, from the previous eruption, from the river or the lake as it was with a Geyser Bolshoi (The Big) after the catastrophe on June 3, 2007. In process of structure filling the water in the bottom of channel and boiler heats up to boiling temperature. Steam formation begins: vapor bubbles in the channel rise up and go to air, bubbles in a boiler accumulate under its arch. Eventually in process of warming up and boiling the steam under the arch of a boiler expands squeezing out water from structure via the channel. This can be considered as a phase of water overflow (quiet expiration of water from the channel ). At last the boiler becomes empty of water, the vapor pressure increases to $\rho g H$ – hydrostatic pressure of water column in the channel. And vapor from a boiler starts to penetrate the channel. Let's consider this moment in more detail.

Let's assume that the steam volume in a boiler increased by the small size $\Delta V$ and became equal to $V_b + \Delta V$. Obviously, it occurred due to the replacement from the channel of the same volume of water: $\Delta V = S \Delta h$, where $\Delta h$ is the corresponding reduction of height of a water column. Respectively, pressure of water in the channel on the border with vapor decreased by $\Delta p_{ch} = \rho g \Delta V / S$. Pressure in a vapor bubble (and accordingly in a boiler) decreased too, but under the adiabatic law (gas conservation law) $pV_b^\gamma = A = const$, where $\gamma$ is the adiabatic curve indicator (for water vapor $\gamma = 1,4$), and this reduction depends on value of $V_b$. At rather large volume of a boiler the pressure drop in a bubble can be much less than the pressure drop of a water column over a steam bubble. This difference of pressure starts the vapor to push out water from the channel like a piston.

Let's translate it into mathematical language. At vapor penetration from boiler to channel the vapor pressure decreases by the value $\Delta p_b$ equal:

$$\Delta p_b = \frac{dp_b}{dV_b}\Delta V = \frac{A\gamma}{V_b^{\gamma+1}}\Delta V \qquad (1)$$

The condition $\Delta p_b < \Delta p_{ch}$ serves as instability criterion. So

$$\rho g / S > A\gamma / V_b^{\gamma+1} \qquad (2)$$

The constant $A$ can be found from a condition of equality of vapor pressure at the bottom of the channel and a vapor pressure in a boiler $\rho g H = A/V_b^{\gamma}$ from where follows: $A = V_b^{\gamma}\rho g H$. Substituting this value in (2) we receive an eruption condition (criterion of instability) or a condition of existence of a geyser:

$$V_b > \gamma H S \qquad (3)$$

The condition (3) means that the instability of water-vapor system and the corresponding pushing out of water by vapor can come only if the volume of a boiler exceeds volume of the channel of a geyser at least in 1,4 times. Obviously if $V_b \gg HS$ the vapor pressure in a boiler will remain practically equal to $\rho g H$ while the pressure of a water column will decrease. Thus water pushing out by the vapor will demonstrate the acceleration until the vapor pressure force is balanced by the channel friction force. This is the beginning of the eruption.

Eruption comes to an end when all water is pushed out from the channel and vapor freely comes to surface. The boiler is released from vapor, pressure and temperature in it fall, there is a steaming phase, and the geyser returns to original state. Eruption can be interrupted owing to the loss of energy of the vapor making work on pushing out of water column from the channel. This loss can lead to decrease in temperature of vapor and to its condensation.

It should be noted that the system of cavities or cracks connected together and having contact to the channel can act as a boiler.

If the volume of a boiler slightly exceeds the channel volume, the eruption can interrupt when a vapor pressure in the channel become equal to hydrostatic pressure of the water remained in the channel. If the height of a column of this water is designated by $h$, the condition of interruption of geyser eruption (a condition of equality of pressure) will look like:

$$h = \frac{V_b^\gamma H}{[V_b + (H-h)S]^\gamma} \qquad (4)$$

**3. Model application to real conditions and discussion.**

Assumptions of existence in structure of a geyser of special chamber became long ago and in many works. Probably the unique work (Lloyd, 1975) contains data on the real structure of the non-operating geyser. There were data received in 1907 about extinct New-Zealand geyser Ta Vera. Its channel depth was 4 meters and one person squeezes inside the geyser (i.e. diameter of the channel was about 0,5 meters). In the bottom of the channel he discover a connection with lateral chamber 3,6 x 2,7 m by size.

The empirical data not contradicting our "boiler" model were received by A.Belousov (the oral message) at research by means of a video camera in channels of geysers Velikan and Bolshoi in the Valley of Geysers. So in a geyser Bolshoi on depth of 3 m the narrowing of the channel and its transition to a narrow lateral crack was found. From this crack time to time a superheated vapor gushed out.

The more a volume of a boiler ($V_b \gg HS$) the more an overflow is prolonged (volume of the effused water doesn't exceed boiler volume) and the more sharp becomes the eruption. If the volume of a boiler is commensurable with the channel volume ($V > \gamma HS$) and if the water arriving in geyser structure suffices only for boiler filling, eruption without overflow is possible (geysers Changeable, Double, The Bastion). An instability development in this case is weaker and eruption can be incomplete: the water will splash out on a small height without a steaming.

If an instability isn't present a vapor stably comes out by the channel – there is a boiling source which can operate in a periodic mode (a periodic devastation of boiler and its filling with water). If water gets to a boiler via the narrow channel and a boiler is constantly filled by the vapor, conditions for instability can be carry out at once: the geyser will "spit out" water with the period in some seconds (the geyser The Torch was such) or represent almost constantly spouting fountain (geysers Averyev, the New Fountain).

Let's note also that the geyser won't be erupting if it isn't provided by free overflow from a cone or crater. Water evacuation from the channel owing to vapor pressing from below and the corresponding fall of the water column pressure are necessary conditions of instability. If the channel of a geyser opens at the bottom of a rather big pool (as geysers The Bath and The Grotto are) it is difficult to the "water-vapor" system to reach a critical condition. Really (Fig. 2) if the vapor volume of $\Delta V$ is pushed out into the channel the water pressure drop in a standard geyser will be $\Delta p = \rho g \Delta V / S$ and in a geyser-bath respectively $\Delta p = \rho g \Delta V / S_1$, i.e. in $S_1 / S$ time it is less. At $S_1 \gg S$ the instability can't develop.

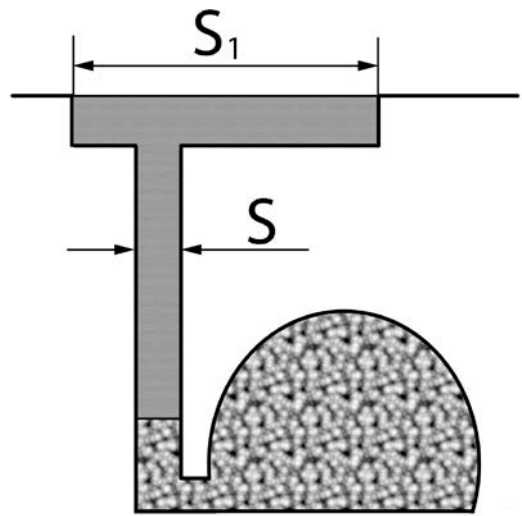

Fig. 2 The geyser-bath structure has the section of a bath considerably exceeding the section of a channel. Eruption don't occur, or it is weakened.

The geyser Bath in the Valley of Geysers has a pool with the cross-section about 3x2 m and depth 0,5 – 0,7 m. The channel of a geyser opens on the bottom of a pool sideways and has visually diameter no more than 0,5 m. Each 5-7 minutes the breaker of the foamed water escapes from the channel and is splashed out on height 0,3 – 0,5 m. Before this splash the water level in a pool increases as though something pushes it in depth. In this geyser $S_1/S \approx 25$. That is the instability mechanism apparently works but it suffices only for a certain portion of vapor which comes upward.

The channel of geyser Grotto with diameter about one meter apparently has his top aperture in a cave which is filled with the small lake-pool (the size about 10x3 m), this pool has walls of geyserite mineral in height about 20 sm. Each 15-20 minutes from depths of the channel the vapor breaker rises upward increasing a water level in a pool and plums of excess water flow down the Big Stained-glass slope. Eruption doesn't occur because a ratio is less than 25. In 1991 Vitaly Nikolayenko, the ranger of the Kronotsky reserve, carried out an experiment having disassembled geyserite wall of the pool. The water left a pool, the channel of the geyser was discharged and the Grotto started to erupt (Nechayev, 2007). Obviously lack of a pool over the channel created a standard situation for a geyser. Eruptions were so strong that in order to avoid destructions on the uniquely beautiful Big Stained-glass slope the reserve administration decided to "cement" a break in a pool wall of the Grotto. And the geyser ceased to erupt once again.

The ejections of water which are not coming to eruption ( the so-called "false starts") are characteristic for fountain-type geysers with rather big diameter of the crater: Velican (Giant), Bolshoi, First-born. When superheated vapor from a boiler arrives to the channel and pushes out a part of water out of channel borders, sharp expansion of a bubble begins as the pressure of water round it (from above and on each side) appears less than a vapor pressure. The bubble pushes water before itself and it is thrown on some height over a crater. However if water temperature is insufficiently high, the bubble can condense and collapse and eruption will interrupt at the beginning. The part of water from the channel returns again in a boiler. Such is one of "false start" mechanisms: begun but not developed eruption. The abovementioned decrease in vapor temperature owing to its work can be the other mechanism. «False starts» effectively mix water in the wide crater of a geyser, warm up total height of the column and approach the general eruption. It can quite occur with a boiling up of superheated water which is pushed by the vapor in the area of low pressure (the classical mechanism of geyser eruption, Droznin, 1982).

Let's consider now a grandiose though cruel experiment which was carried out over Kamchatka geysers by the Nature (Nechaev, 2007). On June 3, 2007 in the Valley of Geysers the powerful mudflow descended. Multi meter thickness of rocks partitioned off the river Geyzernaya and formed the lake by depth to 20 meters. As a result geysers Maly (Small) and Bolshoi were founded under water. Maly – on depth of 15 meters, Bolshoi – on depth of 2 meters. Naturally, geysers couldn't erupt because over them there was a layer of water ( $S \approx \infty$ ) and conditions of instability couldn't be carried out. Over a geyser Bolshoi the spot of more smooth water testifies that the geyser is alive: it warms up and maybe pushes out from channel a hot water. The river slowly pierced a dam, lake level slowly fell, and in three months it decreases approximately by 2 meters. As a result the crater of Bolshoi was risen over water. At this moment the geyser started to erupt! Nature of eruptions was former, only the period almost became three times shorter: from 1,5 hours decreased till 30-40 minutes. After the geyser was erupted the lake intensively filled it by the water through lowering of edges of the crater. The geyser was filled again to edges less than in one minute. Obviously the period decreased. Further there was a standard process with warming and boiling water up to eruption. It is clear that tons of water with temperature of 26 °C filled in by the lake can't be heated up to boiling temperature for half an hour. Thus this natural experiment disproved all classical mechanisms of eruption connected with an overheat and volume boiling up of water in the channel. As to our "boiler" mechanism it should work as the structure of a geyser remained intact (including a boiler and temperature of underground rocks) and all conditions for eruption are carried out as soon as the edges of geyser crater were risen over water level.

The proposed mechanism allows to explain the behavior of some "non-standard" geysers, for example, eruption of geyser Beehive in Yellowstone National Park. This geyser has two surface aperture of channels: small and big, one of which (small) acts as "presage". In 20 minutes prior to the beginning of the main eruption a small fountain of "presage" – the Beehive indicator – starts to spurt. Our theory helps to explain this "difficult" case. Really it is enough to assume that the boiler of a geyser is connected to surface with two channels – wide and narrow – how it is shown in the Fig. 3. The steam accumulating below the arch of a boiler firstly reach the entrance of narrow channel ( position 1, Fig. 3).

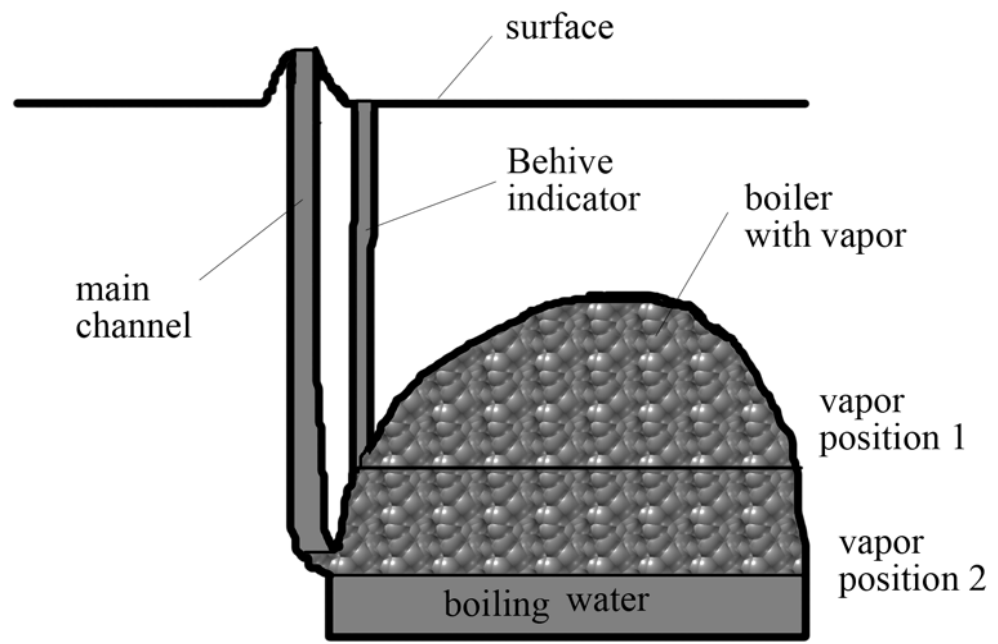

Fig. 3 The supposed structure of Beehive Geyser (Yellowstone National Park)

The instability condition (3) is satisfied and "presage" eruption begins. The length of channels is rather large therefore eruption via narrow channel doesn't manage to pass to a steaming stage (time of pushing out of water in the narrow channel owing to its small section exceeds 20 minutes). Internal apertures of channels are located close from each other, as well as the external. During "presage" eruption the vapor pressure in a boiler continues to increase as the narrow channel is filled with erupting water and steam doesn't come to surface. Through certain time (20 minutes) steam in a boiler reaches an entrance of the big channel ( position 2, Fig. 3). Criterion (3) for the wide channel is satisfied too and the main eruption begins.

According to the instability mechanism and criterion (3) the depth of a channel *H* is a distance from the top of geyser cone to the point of connection of the channel to boiler. *S* is the section of the channel in its top part. In case of geysers-fountains this is the section of a crater. A narrowing cone facilitates and strengthens the eruption, an extending crater complicates and weakens it.

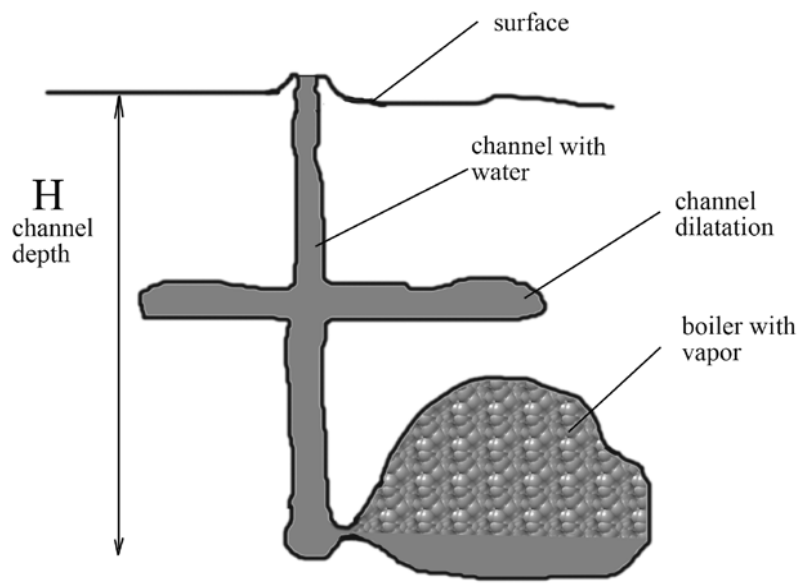

Fig. 4 The supposed structure of geyser with channel dilatation.

If the channel has dilatation on some depth (Fig.4) an instability condition (3) will remain the same independently of form of this dilatation if the sections of channel at its top and bottom are identical. Local dilatation of a channel can influence eruption process, its duration, amount of the erupted water and respectively a period of geyser as time of filling with water of the subterranean plumbing can increase. Dilatation can stop geyser eruption when the steam which is pushing out water in the lower part of the channel reach the dilatation. To penetrate into the top part of the channel steam should release a dilatation cavity from water. The necessary increase in steam volume (it is equal to the dilatation volume) can cause its cooling and interrupt eruption. In this case, obviously, the phase of steaming should be excluded. The increased periods of geysers of Yellowstone (hours and tens hours), the process of eruptions without the subsequent steaming, large volumes of the erupted water can be explained not only by considerable depth of boiler location but also by existence of similar dilatations of channels.

To demonstrate real existence of described mechanism of instability a simple physical experiment can be done. For this purpose the plastic bottle (by volume 1,5 – 2,0 liters) with a cover screwing up densely and a silicone or rubber tubule in diameter of 5-7 mm and near 1m length is sufficient. In a cover the corresponding aperture is bored through into which the tubule end is densely inserted. It is necessary to coat with glue the place of contact of a tubule and a cover to provide tightness of connection. Then the cover with a tubule is wound on a bottle (Fig.5). Functionality of the device is checked by air inflation in a bottle through a tubule: the cover shouldn't let the air out. The bottle simulates a boiler. The tubule is an analog of the channel of geyser, its length is the channel depth. The effect of eruption is shown as follows. The bottle is put vertically (Fig. 5), a tubule lifted over a bottle.

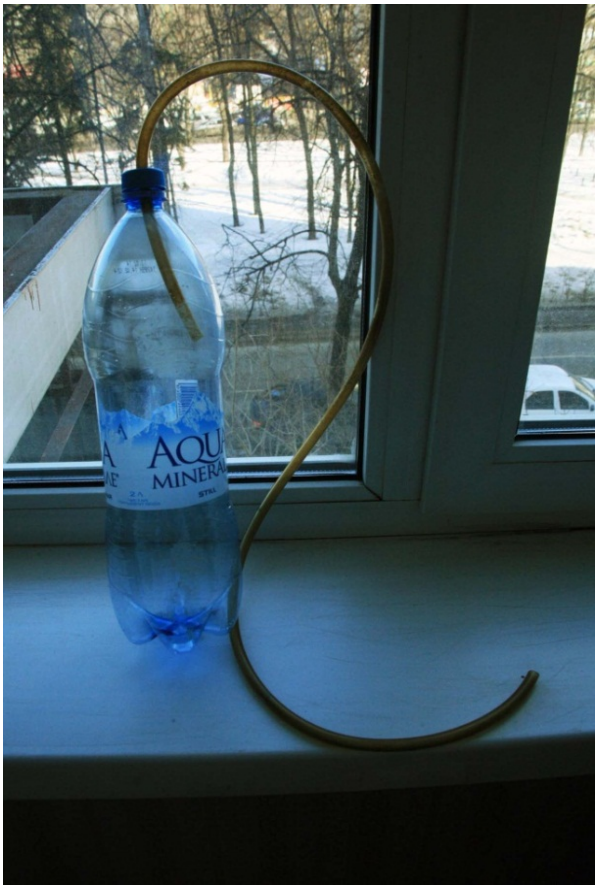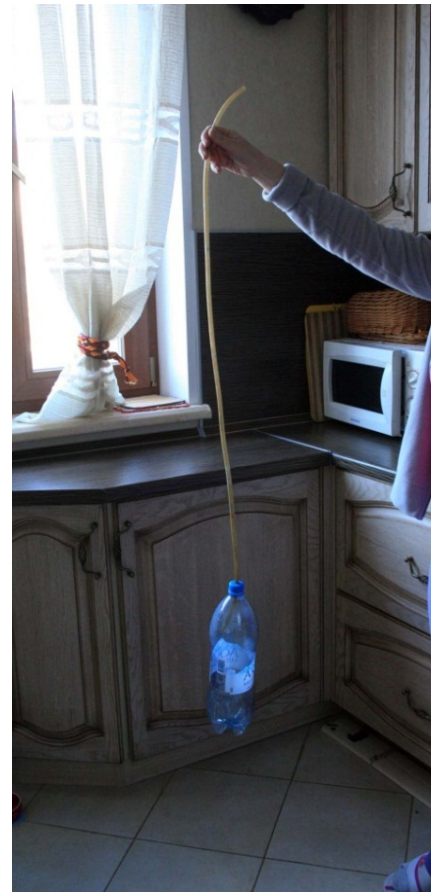

Fig.5 The "device" for demonstration of geyser eruption mechanism.

Water with a thin stream from any vessel is filled in. Water through a tubule flows down in a bottle, thus water should fill all inner diameter of a tubule so that air from a bottle didn't come outside. Pressure of air in a bottle increases and counterbalances hydrostatic pressure of water in a tubule. Water ceases to flow in a bottle and practically during this moment water eruption from a tubule begins. Really the bottle volume is much more than the tubule volume, the criterion (3) is obviously executed so the slightest concussion of a tubule deduces system from a condition of unstable balance therefore all water erupts up with small fountain.

## Conclusions

It is shown that in geyser model with a lateral subterranean chamber connected with channel the new mechanism of instability leading to water eruption is available. The theoretical ratio between parameters of the geyser structure necessary for this instability is defined. The possibility of real existence of this mechanism is discussed basing on the field observations in the Valley of Geysers of Kamchatka.